\documentclass[11pt]{article}
\usepackage{moriond,epsfig}

\def\be{\begin{equation}}
\def\ee{\end{equation}}
\def\bea{\begin{eqnarray}}
\def\eea{\end{eqnarray}}

\begin{document}
%%% T.H.Kress: 2 cm space instead of 4 cm !
%%%\vspace*{4cm}
\vspace*{4cm}
%%%
\title{PARTICLE CORRELATIONS IN Z AND WW EVENTS}

\author{T.H. KRESS}

\address{University of California, Riverside and OPAL Collaboration \\
E-mail: Thomas.Kress@cern.ch}

\maketitle

\abstracts{
Important information about the dynamics of hadron production can be obtained
by the study of particle correlations.
More than 16 million hadronic $\mathrm{Z^{0}}$ decays and several thousand 
$\mathrm{W^{+}W^{-}}$ events have been recorded from the four LEP collaborations
between 1989 and 2000. 
Recently, in $\mathrm{Z^{0}}$ decays, new results of Bose--Einstein correlations
in pairs of pions and Fermi--Dirac correlations for antiproton pairs were reported.
In fully--hadronic $\mathrm{W^{+}W^{-}}$ decays particle correlations were
used to study whether the two W bosons decay independently.}

\section{Introduction}
From 1989 until 1995 the LEP collider operated at centre--of--mass energies 
around 91\,GeV which allowed each of the four experiments to record more than 
four million hadronic $\mathrm{Z^{0}}$ decays.
After the collider energy had been increased above the WW threshold, each 
experiment recorded about ten thousand $\mathrm{W^{+}W^{-}}$ events 
until the end of LEP in 2000.
A hadronic decay of a Z or W boson leads to some dozen particles in the
final state, mostly charged pions and photons from the decay of the 
$\mathrm{\pi^{0}}$ mesons, but also, to a lesser extent, to kaons, protons and 
$\Lambda$--hyperons, which allows to study particle correlations in detail and
thus to get important information about the hadron production mechanism.
%%%

%%%
\noindent
Bose--Einstein correlations (BEC) between identical bosons are well 
established in high energy physics experiments and are often 
considered to be equivalent to the Hanbury Brown \& Twiss~\cite{bib:hbt} 
(HBT) effect in astronomy, describing the interference of photons emitted 
incoherently. 
An alternative approach was proposed by Andersson et al.~\cite{bib:lund},
taking into acount the dynamics of hadron formation in a coherent production
process within the framework of the Lund string model, related to the
symmetrisation of the quantum--mechanical amplitude.
%%%

%%%
\noindent
Bose--Einstein correlations lead to an enhanced production of pairs of identical 
bosons with a small four--momentum difference 
$\mathrm{Q^{2}=-(p_{1}^{\mu}-p_{2}^{\mu})^{2}}$.
Traditionally, BEC are studied using a two--particle correlation function
$\mathrm{C(p_{1},p_{2}) = \rho_{2}(p_{1},p_{2})/\tilde{\rho}_{2}(p_{1},p_{2})}$, 
where $\mathrm{\rho_{2}}$ and $\mathrm{\tilde{\rho}_{2}}$ are the two--particle 
densities with and without BEC, respectively.
For the construction of the reference sample $\mathrm{\tilde{\rho}_{2}}$ 
frequently a MC model without BEC is used.   
Following the pioneering analysis of Goldhaber, Goldhaber, Lee and 
Pais~\cite{bib:gglp} (GGLP), a correlation 
function of type $\mathrm{C(Q) = 1 + \lambda\exp(-Q^{2}R^{2})}$ is often used 
to yield a value for R, which is interpreted as the emitter radius. 
The factor $\mathrm{\lambda}$ measures the strength of the BEC effect but 
sometimes absorbs also experimental inpurities.
%%%

%%%
\section{Particle Correlations in $\mathbf{Z^{0}}$ Decays}

\subsection{Bose--Einstein Correlations in Pairs of Pions}
\begin{figure}[h]
\epsfxsize=18.5pc \epsfbox{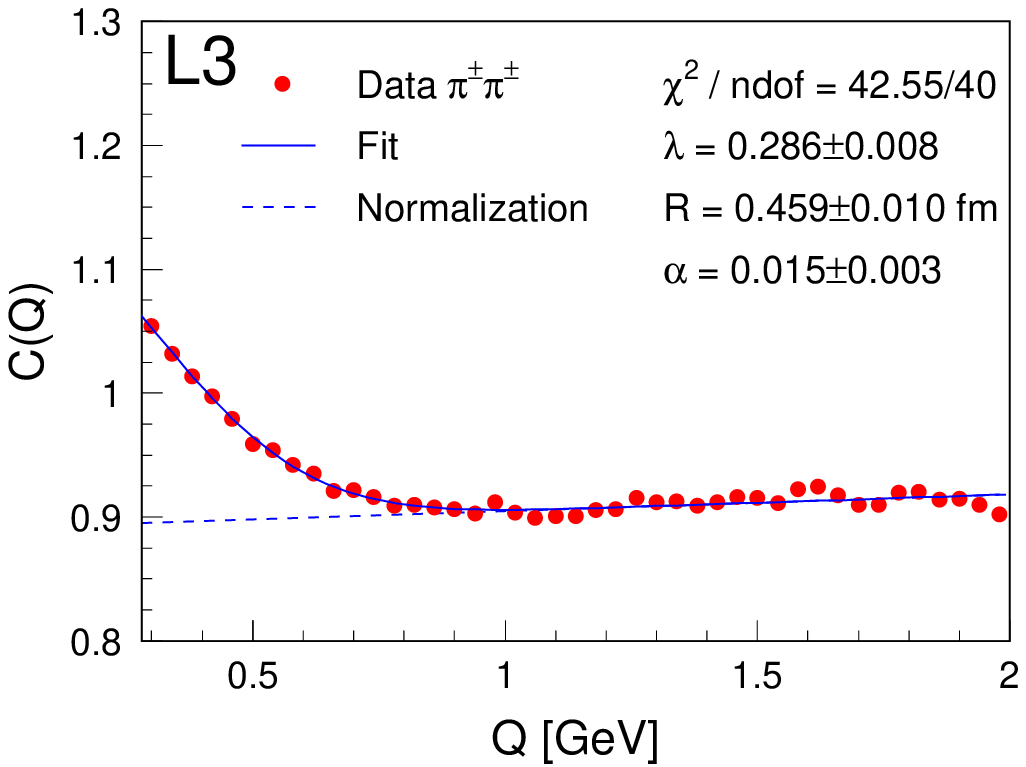}
\epsfxsize=18.5pc \epsfbox{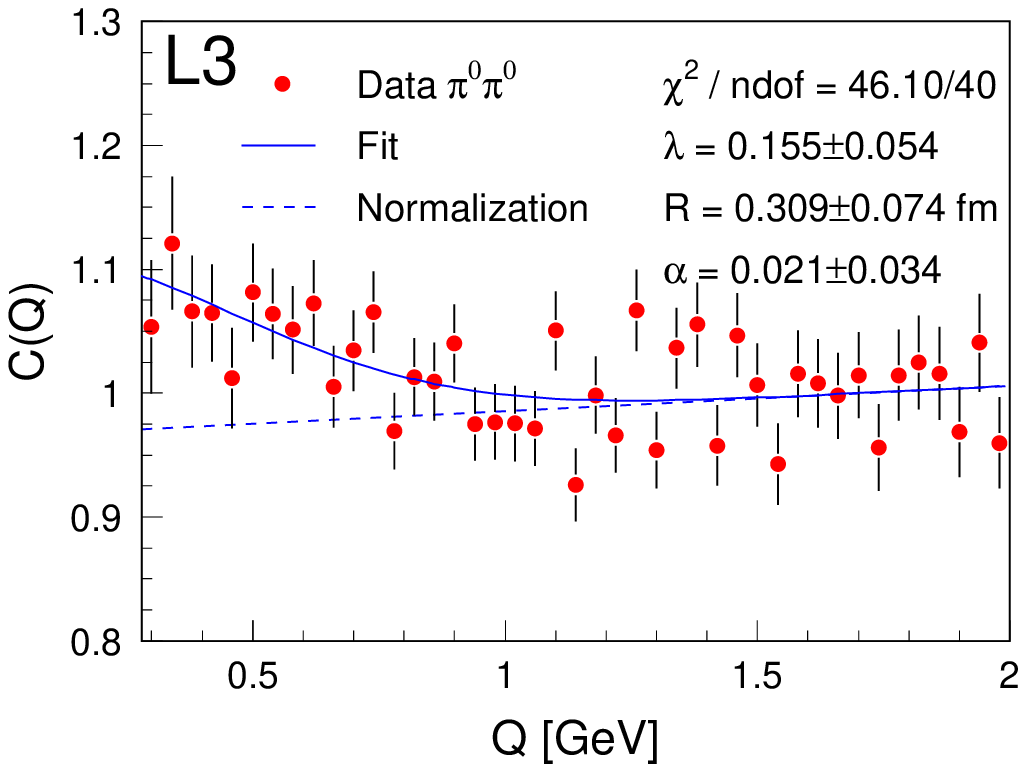}
\caption{Correlation function C(Q) for pairs of charged and neutral 
pions.\label{fig:pion}}
\end{figure}

Fig.~\ref{fig:pion} shows recent L3 measurements~\cite{bib:l3_pion} of the correlation
function for charged (left) and neutral (right) pion pairs, using a MC without BEC as
the reference samples.
%%%

%%%
%%%

%%%
\noindent
For both the charged and neutral pion pairs an enhancement at low Q is 
visible. 
Using a GGLP type of parametrisation, the obtained source radius R
for neutral pions is tending to be smaller than for
charged pions, as qualitatively expected in the Lund string
model~\cite{bib:zalewski}.
%%%

\subsection{Fermi--Dirac Correlations in Pairs of Antiprotons}
It has been proposed~\cite{bib:alexander_fd} to extract an emitter dimension 
for pairs of equal baryons by utilising the Fermi--Dirac exclusion principle. 
The correlation function can be parametrised by an equation similar to the 
GGLP parametrisation with the plus sign replaced by a minus sign.
Antisymmetrising the total wave function yields four states, three of which 
are antisymmetric in space and symmetric in spin. Thus, for an incoherent source, 
C(Q) should decrease to a value 1/2 in the limit $\mathrm{Q \rightarrow 0}$.
%%%

\vspace*{0.5mm}

%%%
\begin{figure}[h]
\epsfxsize=18.0pc \epsfysize=12.0pc \epsfbox{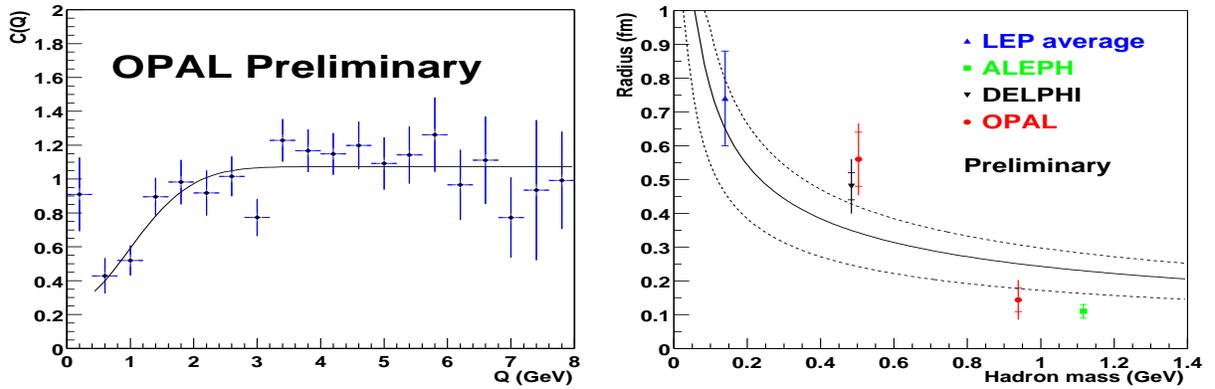}
\hspace*{0.5pc}
\epsfxsize=18.5pc \epsfysize=12.0pc \epsfbox{letts_7.epsi}
\caption{Correlation function for anti--protons (left) and emitter radius as function 
         of the hadron mass (right).\label{fig:proton/Rmass}}
\end{figure}
%%%

%%%
\noindent
As a preliminary result~\cite{bib:opal_proton}, the OPAL collaboration reported a depletion
of anti--proton pairs at low Q, as shown on the left--hand side of Fig.~\ref{fig:proton/Rmass}.
%%%

%%%
\subsection{Dependence of the Emitter Radius on the Hadron Mass}
Together with results for the emission radius for pion, kaon and $\Lambda$ pair
correlations, compilated by Alexander et al.~\cite{bib:alexander_Rmass}, the radius R 
for pairs of antiprotons is shown as a function of the hadron mass on the right--hand 
side of Fig.~\ref{fig:proton/Rmass}.
The observed hierachy $R_{\pi} > R_{K} > R_{\bar{p},\Lambda}$ can be explained 
qualitatively by a model~\cite{bib:alexander_Rmass} based on the Heisenberg uncertainty 
principles and an approach~\cite{bib:bialas_Rmass} taking into account the strong
correlation between space/time $x^{\mu}$ and momentum/energy $p^{\mu}$ of the particle 
in the hadron production process.

\section{Inter--WW Bose--Einstein Correlations}
In $\mathrm{W^{+}W^{-} \rightarrow q\bar{q}q\bar{q}}$ events at LEP,
the products of the W decays have in general a significant 
space--time overlap as the separation of the their decay vertices
is small compared to characteristic hadronic distance scales.
The W boson mass, a fundamental parameter in the Standard Model, 
is determined from the corresponding jet masses and could
potentially be biased~\cite{bib:rabbertz} if Bose--Einstein correlations 
between the decay products of the two W bosons exist. 
A robust framework to test the presence of such inter--WW BEC
was proposed by Chekanov et al.~\cite{bib:cdwk}. If the $\mathrm{W^{-}}$ and 
$\mathrm{W^{+}}$ decay independently, then $\mathrm{\Delta\rho}$(Q)$=$0
for all Q with the test distribution $\mathrm{\Delta\rho}$ defined as:
%%%
\begin{displaymath}
\Delta\rho = 
\rho_{2}^{WW\rightarrow 4q} - 2\cdot \rho_{2}^{W\rightarrow2q} - \rho_{2}^{WW_{mix}},
\end{displaymath}
%%%
with the two--particle densities $\mathrm{\rho_{2}^{WW\rightarrow 4q}}$
determined by the 
$\mathrm{W^{+}W^{-} \rightarrow q\bar{q}q\bar{q}}$ sample, 
$\mathrm{\rho_{2}^{W\rightarrow2q}}$ by the hadronic part of
semileptonic $\mathrm{W^{+}W^{-} \rightarrow q\bar{q} l\bar{\nu}}$ events 
and $\mathrm{\rho_{2}^{WW_{mix}}}$
from events build from two independent semileptonic events without
the leptonic parts and combining only particles originating from
different W's.
%%%

%%%
\begin{figure}[h]
\epsfxsize=18.0pc \epsfysize=15.2pc \epsfbox{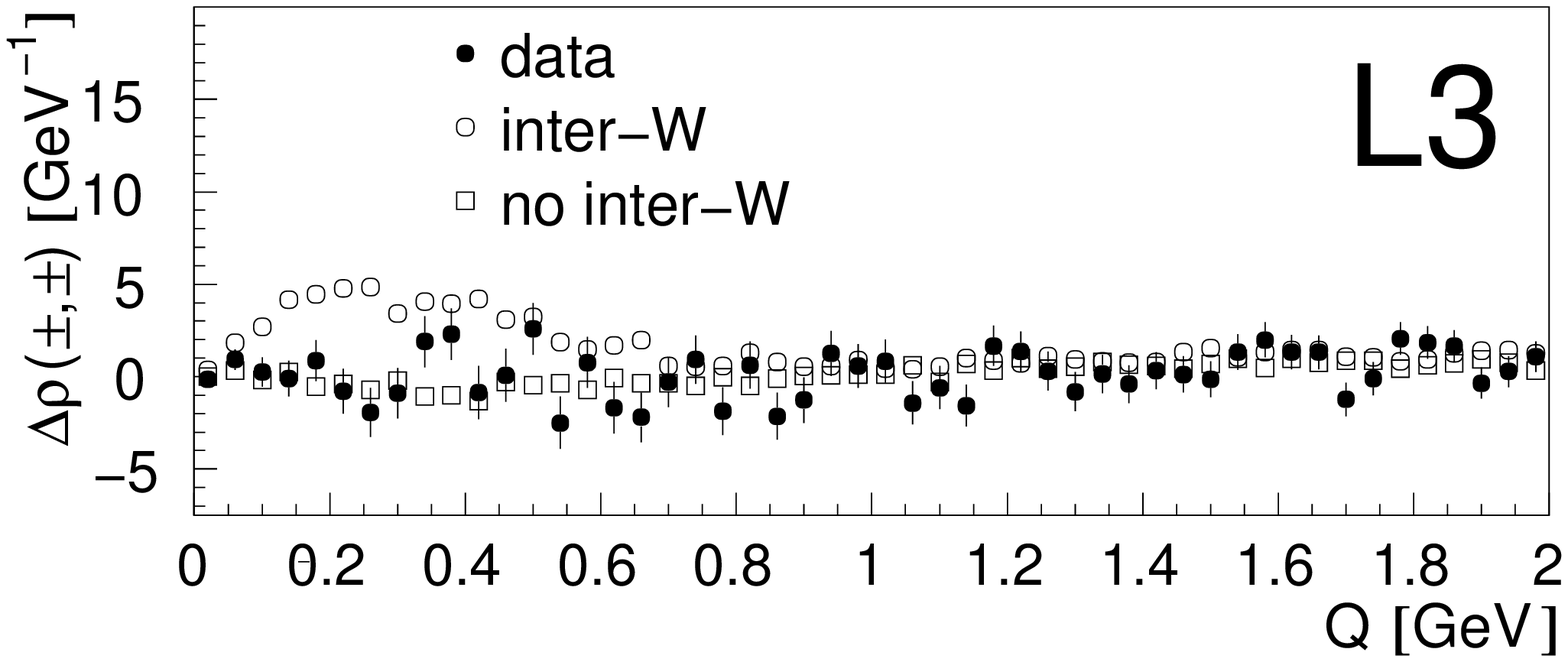}
\hspace*{1.0pc}
\epsfxsize=18.0pc \epsfysize=15.0pc \epsfbox{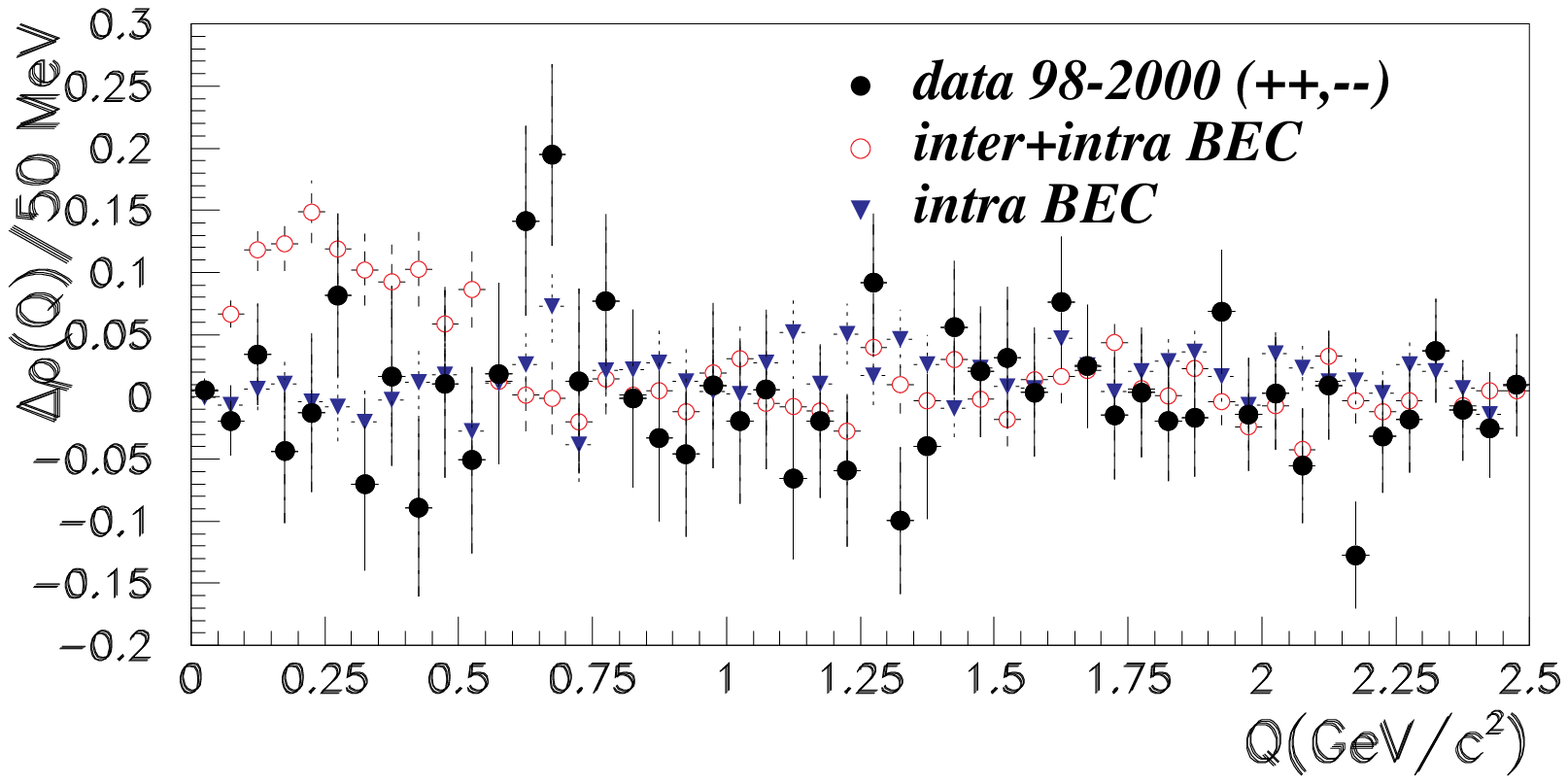}
\caption{Preliminary $\Delta\rho$ distributions from L3 (left) and DELPHI (right) 
compared with MC models.\label{fig:Drho}}
\end{figure}
%%%

%%%
\noindent
In Fig.~\ref{fig:Drho}, the L3 collaboration~\cite{bib:l3_ww} compares the 
$\mathrm{\Delta\rho}$ distribution obtained from the data with two scenarios of the 
PYTHIA/PYBOEI~\cite{bib:pyboei} Monte Carlo model with BEC. 
The inter--WW BEC in the Monte Carlo model can be seen as an enhancement
of like--sign pairs in the low Q region. 
The data are consistent with no inter--WW correlations and the 
Monte Carlo, which describes BEC between particles from different
W bosons in the same way (i.e. same R, $\lambda$) as the correlations within the 
same W, is strongly disfavoured.
The same is true for similar results from DELPHI~\cite{bib:delphi_ww},
as shown on the right--hand side of Fig.~\ref{fig:Drho}.
Both results are preliminary. 
%%%

%%%
\noindent
In the Lund model, Bose--Einstein correlations arise when identical 
bosons are produced close to each other within the same string.
Because of the strong correlation of production space--time and momentum of the
hadron, the measured R is interpreted as the distance in the string where
the momentum spectra of the particles still overlap. 
In the absence of colour reconnection effects~\cite{bib:rabbertz}, particles
from different W bosons are not produced in the same string.  
However, in addition to the coherent correlations inside a string,
a second correlation effect of an incoherent HBT type could be present.
An analysis~\cite{bib:todorova} of the hadron formation 
within the Lund model shows that the
space--time distance of the production vertices for pairs of particles 
from different strings is of the order of several fm.
For such large distances any remaining inter--WW BEC effect would manifest
itself only at very low Q values which are hard to exploit with the limited 
statistics of WW events at LEP~\cite{bib:dewolf}.

\section{Summary}
Bose--Einstein correlations in high energy physics have been studied extensively
for more than 40 years. 
It seems that we have now come to a better understanding of the effect by taking
into account the dynamics of hadron production.
%%%

%%%
\noindent
Since no firm theory exists to describe the hadronisation phase, we rely on 
phenomenological models. To test such models, the study of particle correlations 
provides us with details complementary to those obtained from global event 
properties and single--particle distributions.
%%%

%%%

\end{document}